\documentstyle[11pt,moriond,epsf]{article} 

\begin{document}

\vspace*{4cm}
\title{CP VIOLATION IN $ K^0 $ DECAYS}

\author{ Bruce Winstein }

\address{The University of Chicago, 5640 South Ellis Avenue, \\
Chicago, IL 60637, U.S.A.}

\maketitle\abstracts{In this lecture, I will review recent results on
the study of ``direct'' CP violation, where the violation  occurs in
a particle decay rather than in particle-antiparticle mixing.
Brief comments on rare decays, particularly those with a CP reach,
will be made.  Finally, I will indicate how the community
would like to see the field evolve.} 

\section{MOTIVATION} 

	There are a few reasons why pursuit of experimental studies
of CP violation are important.  The first is that we only have one
established effect.  There are many manifestations, all in the
decays of the neutral kaon:  $ 2 \pi $, $ \pi l \nu $, $ \pi \pi 
\gamma $, and, most recently\footnote{Observation of CP Violation in 
$K_L \rightarrow \pi^+ \pi^- e^+ e^- $ Decays, KTeV Collaboration, 
hep-ex 9908020}
in $ \pi \pi ee $.  But these are all traceable to the $ \epsilon $
impurity in the $ K_L $ state:  the sizes of the asymmetries are
precisely predicted and this effect is dubbed ``indirect''
in that it arises from a (presumably) second order CP- and T-
violating interaction in the K-K transition.

	The effect is also of cosmological significance.  We live in
a world of matter: we find no anti-planets, no significant
component of anti-matter in extra-galactic cosmic rays, and no
evidence for anti-galaxies.  In our galaxy, there are
approximately $ 10^{69} $ protons and $ 10^{79} $ photons whereas,
if we extrapolate this volume back to a time of $ 10^{-6} $ sec
from the big-bang, its temperature was about $ 10^{13} $ degrees
and it would have contained $ 10^{79} $ protons, anti-protons,
and gammas.  Nevertheless, it is thought that the baryon
asymmetry was present even at this early time so the question
remains as to its origin, and whether the CP violation we see
in the neutral kaon decays is at all connected.

	The standard model ``explanation'' of the effect (in the
kaon system) is compelling and it predicts a direct effect in K
decays as well as larger effects in the mixing and decays of B
mesons.  Now, we have one effect and one parameter ($ \delta $,
the phase in the CKM matrix, or $ \eta $, the Wolfenstein
parameter):  we badly need more effects, both within the K and B
systems, to see if the model is consistent.  Indeed the future of
K decay experiments is aimed at such studies as we will see later.

\section{MEASURING $ \epsilon^{\prime}/\epsilon $}

	The best avenue to see direct CP violation is the study
of $ \epsilon^{\prime}/\epsilon $ in the neutral kaon decays to
two pions.  All four modes need to be studied and the double ratio,

\[ R = \frac{  \Gamma ( K_L \rightarrow 2 \pi^0 ) /
\Gamma ( K_S \rightarrow 2 \pi^0 ) }
{ \Gamma ( K_L \rightarrow
\pi^+ \pi^- ) / \Gamma ( K_S \rightarrow \pi^+ \pi^- ) }, \]

\noindent
gives $ \epsilon^{\prime}/\epsilon $ through the relation
$ \epsilon^{\prime}/\epsilon $ = (1-R)/6.

	The experimental situation in the early 1990's led to three
new efforts to further search for direct CP violation, these
being KTeV at Fermilab (an evolution of the E731 experiment),
NA48 at CERN (an evolution of the NA31 experiment), and KLOE at
DA$ \Phi $NE, an entirely new effort.  At the present time,
there are first results from both KTeV and NA48, and KLOE is
just beginning to take data.

	For the present generation of experiments, there are
certain common features:

\begin{itemize}

\item	collection of all four modes simultaneously by means of
two beams, one with $ K_L $ decays and the other with $ K_S $
\item	use of high-precision electromagnetic calorimetry for
excellent reconstruction of the 2$ \pi^0 $ decays
\item	precision magnetic spectrometry.

\end{itemize}

\noindent
This leads to relatively low and understandable backgrounds as well.

	The areas where there remain differences are the following:
\begin{itemize}

\item	how do the experiments make $ K_S $?
\item	how do the experiments account for the acceptance difference
(arising from the differing lifetimes) between $ K_S $ and $ K_L $?
\item	is the analysis done blind?
\end{itemize}

\section{KTEV}

	The first result from KTeV has been published\footnote{A.
Alavi-Harati et al., Phys. Rev. Lett. \textbf{83}, 22 (1999)}
so here I will just briefly discuss some
of its key features.  The KTeV apparatus is shown schematically in
Figure 1.

\begin{figure}[!ht]
\centering
\epsfysize=3.5in   
\hspace*{0in}
\epsffile{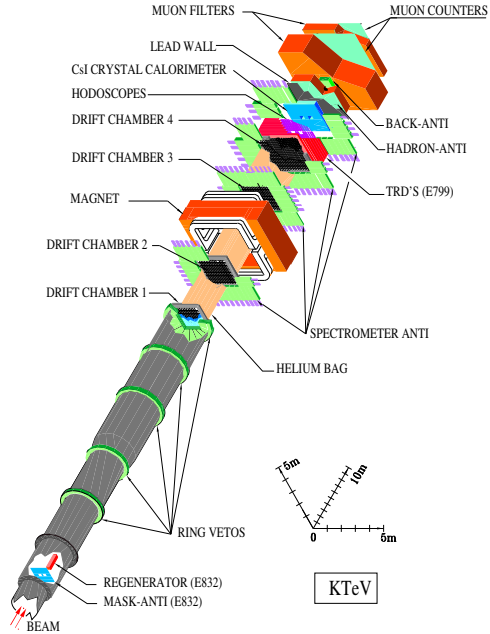}
\caption{Schematic of the KTeV apparatus.}
\end{figure}

	KTeV uses a regenerator to derive its $ K_S $ beam.  Shown
in Figure 2 is the reconstructed decay distribution for 
$ \pi^+ \pi^- $ decays downstream of the regenerator, for events
with momentum between 30 and 35 GeV/c. In this bin, the $ K_S $
lifetime is only about 1.7 m so that, to normalize the 
$ \epsilon^{\prime}/\epsilon $ measurement, only a few
meters would be needed.  As the $ K_S $ component decays, it
becomes of comparable magnitude to that of the much smaller
$ K_L $ component and their interference is
most noticable, at about 15 m downstream of the target.  Thus
using all the decays downstream of the regenerator gives access to
the $ K_S $ lifetime, the mass difference between $ K_S $ and $ K_L $,
and the phase difference between the two decays, this for both
$ \pi^+ \pi^- $ and 2$ \pi^0 $ decays.  And the ability to
understand these distributions gives confidence that the detector
and beam are well understood.

\begin{figure}[!ht]
\centering
\epsfysize=3.5in   
\hspace*{0in}
\epsffile{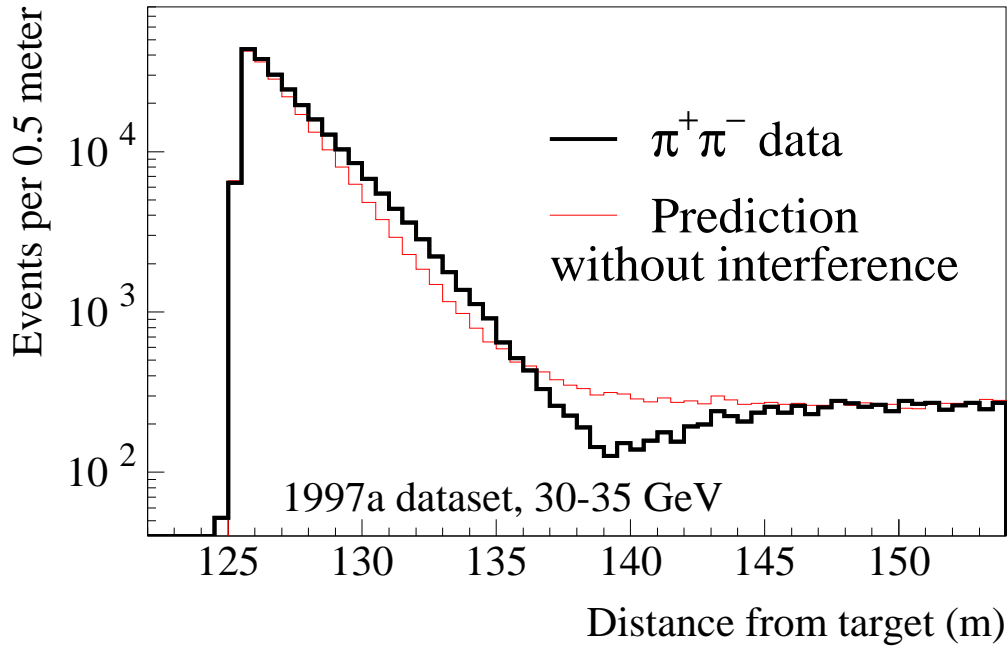}
\caption{Reconstructed $ \pi^+ \pi^- $ data showing interference
downstream of the KTeV regenerator.}
\end{figure}

	There are two drawbacks to the use of a regenerator.  The first
is the ambient rate that it causes in the detector.  Essentially
all the kaons and neutrons incident upon it interact\footnote{The
KTeV regenerator is about 1.8 kaon interaction lengths and about
3.5 neutron interaction lengths.} and therefore
potentially spray unwanted ``accidental'' particles into the
detection elements.  The beam which strikes the regenerator
is filtered and attenuated to enhance the K/n ratio and to reduce
the overall rate.  Nevertheless, the interaction rate in the
regenerator is about 1.5 MHz and this source contributes the
majority of the ambient rate in the detectors.

	This turns out to not be a major problem for several reasons.
The first is that the regenerator is fully active, consisting of 85
separate scintillator pieces each read out by two photomultipliers.
The FNAL beam structure is 53 MHz, meaning that approximately 3\%
of the buckets are occupied, with either a neutron or a kaon, or,
rarely, more than one such particle\footnote{It is important that
the intensity from bucket to bucket be uniform so attention must
be paid to monitoring beam microstructure.}.  Whenever there is an
inelastic interaction producing particles that could hit the detector,
it is effectively vetoed by the signals in the regenerator.  So
the activity will be ``out of time.''

	The KTeV detector with the longest latency, and therefore
most sensitive to such activity, is the drift chamber system
where drifts up to approximately 200 ns occur. Out-of-time
activity can interfere with the real track information if
it occurs on the relevant wires.  This effect has been extensively
studied and to a very high order is properly simulated by simply
superimposing accidental events upon monte-carlo generated ``pure''
events.  By such studies, we have determined that the overall
effect of the ambient rate from the regenerator shifts the ratio
of $ K_S $ to $ K_L $ (``single ratios'') by less than 0.001 in
each ($ \pi^+ \pi^- $, 2$ \pi^0 $) case.  There is a loss of events,
at the level of 2\%, due to such accidental activity in the
chambers,  but the loss is quite symmetric between the two beams.
The main reason for this is that the ``spray'' from the
regenerator is broad enough that it effectively is uniform over
the drift chambers; were the ambient particles confined to a
small region about the (regenerator) beam region, the difference
would be greater.

	The second drawback from the use of the regenerator is
incoherent scattering of $ K_S $.  When such scatters produce
extra particles (inelastic events), they are effectively self-vetoed.
But elastic scattering off Carbon nuclei produce negligible recoil
energy so that these events smear out the coherently
regenerated beam and even can ``cross-over'' to the vacuum side.
Figure 3 shows this effect for neutral events at the calorimeter.

\begin{figure}[!ht]
\centering
\epsfysize=3.5in   
\hspace*{0in}
\epsffile{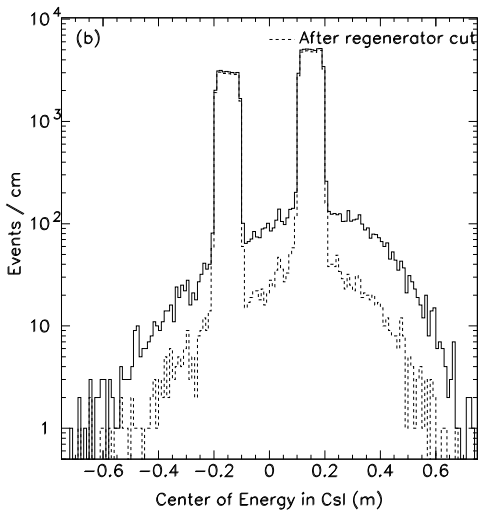}
\vspace*{-0.2in} 
\caption{Reconstructed transverse position of neutral events at the
calorimeter.  The regenerator (which normally moves from side-to-
side) is for this figure always on the right.}
\end{figure}
	
	Fortunately this effect can be directly measured using
the $ \pi^+ \pi^- $ events themselves.  For the angular
distribution of this process is independent of how
the kaon decays.  Hence the distribution is determined with the
charged sample (using the magnetic spectrometer) and this can
be used to accurately give the effects in the neutral decays.  Small
uncertainties in the charged acceptance for this scattered
component, however, give a systematic uncertainty of order
0.0005 in the double ratio.

	We show in the table below the performance of each of the
KTeV systems in comparison to that for the previous generation
of experiments at FNAL.  The calorimeter, made of pure CsI, has
achieved a resolution of 0.75\% averaged over the momentum spectrum.

\begin{table}[htp]\centering
\begin{tabular}{|l|c|c|}
\hline
Parameter & E731/E773/E799 & KTeV \\
\hline
Pressure in decay region & 500 $\mu$Torr & 1 $\mu$Torr \\
$\mu$ flux per proton on target & $4\times10^{-5}$ 
& $2\times10^{-7}$ \\ 
Max. proton flux/spill	&2.0E12	&5.0E12	\\
Calorimeter radiation exposure (E799) &450 rad/E12/week &50 rad/E12/week  \\
\hline 
$\gamma$ energy resolution at 20 GeV/c & 3.5\% & 0.65\%  \\  
calorimeter nonlinearity (3-75 GeV/c) & 10\% & 0.4\% \\
$\pi$/e rejection, calorimeter & $\sim 50$ & $\sim 400$ \\ \hline
Magnetic field ($p_t$ kick) & 200 MeV/c & 400 MeV/c \\
Magnetic field nonuniformity & 5\%  & 1\%  \\
Material in spectrometer (rl) & $\sim 0.87\%$ & $\sim 0.35\%$ \\
Single wire plane resolution & $\sim 85 \mu$m & $\sim 100 \mu$m \\
Track momentum resolution at 20 GeV/c & 0.5\% & 0.25\%  \\  \hline
2$\mu$ efficiency & 82\% & 99\% \\ \hline
Regenerator: Inelastic background in vac. beam & $\sim 1.8\% $ &  
$\sim 0.3\% $ \\ \hline
$\gamma$ veto performance (p.e. / MeV) & 0.02 & 0.2 \\ \hline
$\pi$/e rejection, TRD system & NA & $\sim 150$ \\ \hline
Level 2 clustering-trigger time & 30 $\mu$s & 2 $\mu$s \\
Level 2 tracking-trigger time & 3 $\mu$s & 1.5 $\mu$s \\
Level 2 TRD-trigger time & NA & 1 $\mu$s \\
Level 3 complete event reconstruction & NA & 200k events/spill \\
DAQ output & $\sim20$ MB / spill & $\sim300$ MB/spill \\
Livetime & 0.7 at 0.8E11 p/spill & 0.7 at 3.5E12 p/spill \\ \hline
Offline $\pi^0\pi^0$ mass resolution (MeV/c$^2$) & 5.5 & 1.5 \\
Offline $\pi^+\pi^-$ mass resolution (MeV/c$^2$) & 3.5 & 1.6 \\
\hline
\end{tabular}
\caption{Comparison between E731/E773/E799 and KTeV.}
\label{ta:compare}
\end{table}
	
	The Monte-Carlo simulation in KTeV is most important
because the vertex distributions are so different between
$ K_S $ and $ K_L $.  Accepted $ K_S $ events reconstruct on
average about 5m upstream of $ K_L $  so, to control the ratio of
recorded events to better than 0.001 means that we should understand
any instrumental induced ``slope'' in the acceptance at the level
of about 0.02\% per meter.

	The best means to study the acceptance in the data is
by using the high statistics decay modes that are taken
simultaneously with the 2$ \pi $ decays.  Figure 4 shows the decay
distribution for about 40 million $ Ke_3 $ decays together with the
monte-carlo simulation.  Fitting the ratio of the two to a
linear departure shows that the distrubutions match at the level of
about 0.005\% per meter.  It is relevant to point out that the
lifetime of the $ K_L $ itself would contribute a slope in this
plot of about 0.05\% per meter.  However, the same distribution
for our $ \pi^+ \pi^- $ sample shows a slope three times greater,
about a 2.5 standard deviation effect; accordingly, we use this
larger slope as a measure of the possible systematic uncertainty
associated with acceptance.

\begin{figure}[!ht]
\centering
\epsfysize=3.5in   
\hspace*{0in}
\epsffile{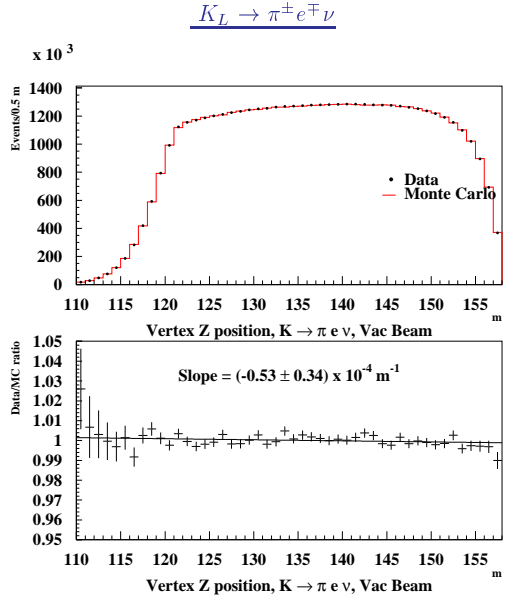}
\caption{Comparison of data and monte-carlo for $ Ke_3 $ events. 
Lower plot shows the ratio of the two.}
\end{figure}

	Making the acceptance corrections, we then look at the
momentum dependence of the ``regeneration'' amplitude for each of
the modes.  This is expected to exhibit a power-law behavour and a
very important check is that the power be the same for the two modes.
We find indeed that the regeneration amplitude can be well
represented by the form $ \mid f-f \mid /k~\alpha~p^{-\alpha} $ and for
the parameter $ \alpha $, we find 0.5890(15) and 0.5884(19) for
the charged and neutral modes respectively.

	We fit for the $ K_S $ lifetime and find consistent results
for each mode.  The KTeV (preliminary) value is 0.8967(7) $ \times
10^{-10} $ s and this is shown in comparison with
other recent measurements in Figure 5.  Again, in fitting for the
mass difference, we again find internally consistent values with
the KTeV value being 0.5280(13) which is shown in the Figure.
Finally, we determine the potentially
CPT violating phase difference between the two CP violating
amplitudes, finding $ \Delta \Phi = 0.09(46)^0 $, as is shown in
Figure 6.  And, at the suggestion of Alan
Kostelecky\footnote{Sensitivity of CPT Tests with Neutral Mesons, V.A.
Kostelecky, Phys. Rev. Lett. \textbf{80}, 1818 (1998).}, we have examined our data for a diurnal variation 
in $ \Phi_{+-} $ as might be predicted in certain CPT and
Lorentz violating interactions, finding no
day-night effect at the level of about 1/3 degree\footnote{To
report a value for $ \Phi_{+-} $, one needs to worry
about certain systematic uncertainties in the phase of the
regeneration amplitude.  But such are irrelevant if one is just 
interested in the \textbf{time dependence} of $ \Phi_{+-} $}.

\begin{figure}[!ht]
\centering
\epsfysize=3.5in   
\hspace*{0in}
\epsffile{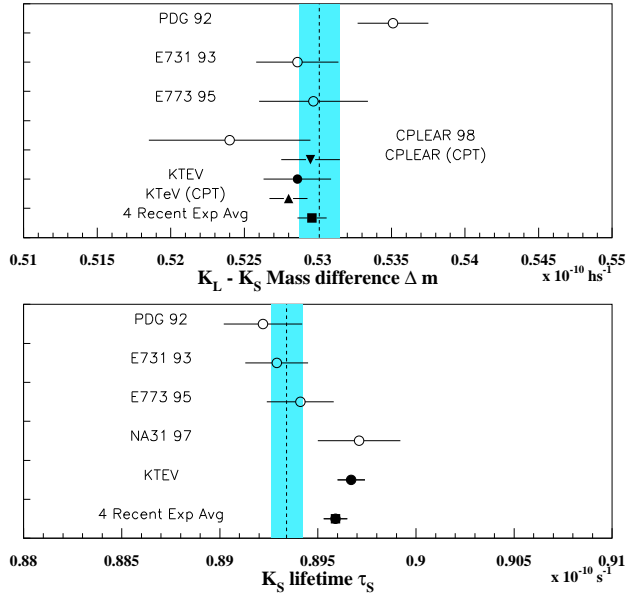}
\vspace*{-0.2in}
\caption{Recent measurements of $ \Delta $m and $ \tau_s $.}
\end{figure}

\begin{figure}[!ht]
\centering
\epsfysize=3.5in   
\hspace*{0in}
\epsffile{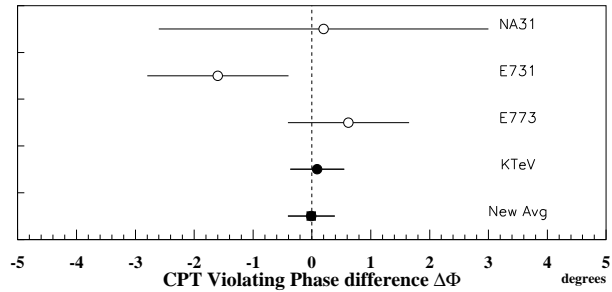}
\vspace*{-1.5in}
\caption{Recent measurements of $ \Delta \phi $.}
\end{figure}

\subsection{Looking at the Answer}

	The $ \epsilon^{\prime}/\epsilon $ analysis is done blind
in KTeV.  

	We use as much as possible the high statistics modes
for study of the detector and to improve our modeling of it,
rather than the 2$ \pi $ modes from which the answer is
calculated.  The acceptance calculations are validated using
$ Ke_3 $ and 3$ \pi^0 $ decays.  Then the 2$ \pi $ samples are
examined and the $ \tau_s $ and $ \Delta m $ parameters are
determined and with high precision are found to be consistent
between the two modes.  A precision CPT test is done, to the
level of 1/2 degree.  And the regeneration powers are found to
be equal with high precision.  Finally, all systematic studies
are completed and the value for each error is tabulated.

	It is only after all of these checks, studies, and
evaluations that the answer is uncovered -- the decision to commit
to a result is made beforehand.

	The result was Re($ \epsilon^{\prime}/\epsilon $) =
(28.0 $ \pm $ 4.1) $ \times 10^{-4} $ where the statistical error was
0.0003 and the systematic 0.00026.  At nearly seven standard
deviations, this establishes direct CP violation.

\section{NA48}

	The NA48 experiment has given its first preliminary
result very recently.  A sketch of their very elegant beam
arrangement is shown in Figure 7.  They too use two beams but the
$ K_S $ beam is derived from a small fraction of the primary
beam that is diverted and then targetted upon a close-by target.
The two beams have different angular divergences but cross at the
detector. (In KTeV, the beams are precisely the same divergence
but are separated at the detector.)

\begin{figure}[!ht]
\centering
\epsfysize=3.5in   
\hspace*{0in}
\epsffile{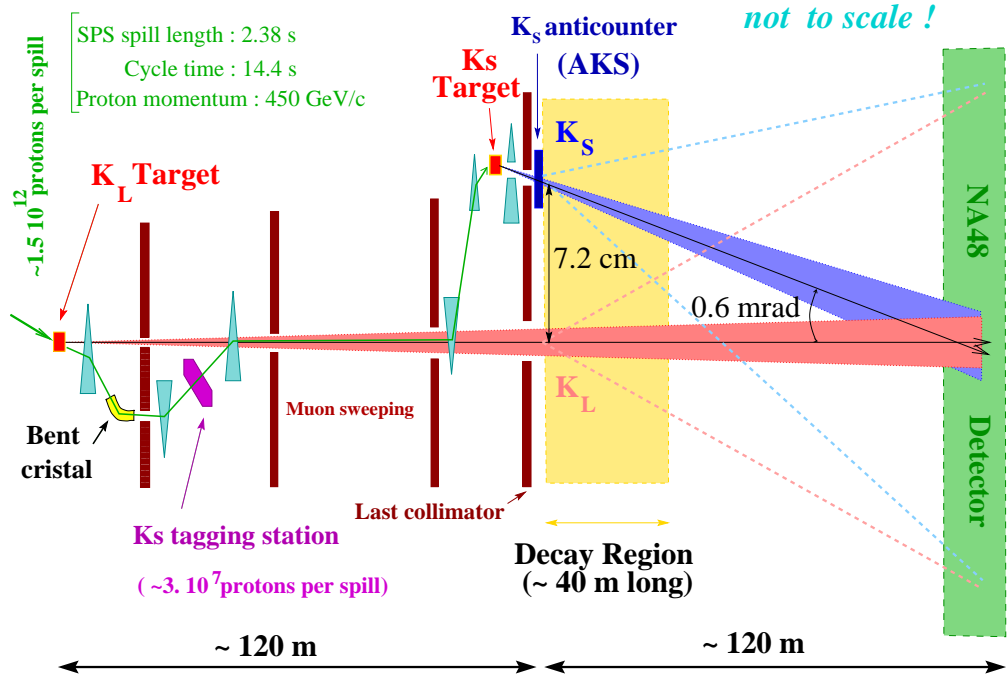}
\caption{Schematic of the NA48 beam system.}
\end{figure}

	The performance of the NA48 calorimenter is excellent,
allowing 0.7\% resolution at high energies.

	In NA31, the $ K_S $ target was serially stepped through
the decay region to approximate $ K_L $ decays; in this way,
acceptance corrections were minimal.  NA48 does not use this
technique but rather has elected to weight their $ K_L $ events
according to the distribution of $ K_S $ events.  This results
in a significant statistical loss but has the advantage that
the corrections that need to be applied, now to the weighted
event ratios, are less than 1\% whereas those for KTeV are of
order 5\%.

	One potential problem that the group has studied
extensively is an energy dependence in the double ratio.  When
fit for a linear slope, a 3 standard deviation departure is
found with the value changing by about 5\% over the span
of energies from 70 to 170 GeV/c.  But extensive studies
of the data, including examining the double ratio in energy
bins beyond their nominal fiducial region, has convinced the
group that this is a statistical fluctuation.

	The preliminary result they report, based upon about
10\% of their anticipated sample, is:
Re($ \epsilon^{\prime}/\epsilon $) = (18.5 $ \pm $ 7.3) $ \times
10^{-4} $ where the systematic error is larger than the statistical
one.  However, most of the systematic error is dominated by
statistics so that, for awhile anyway, the total error will
roughly scale with added data.

\section{THE GRAND AVERAGE}

	The most recent results are shown graphically in 
Figure 8.  The grand average of these last measurements is
(21.2 $ \pm $ 2.8) $ \times 10^{-4} $, with a confidence
level of about 7\%.

\begin{figure}[!ht]
\centering
\epsfysize=3.5in   
\hspace*{0in}
\epsffile{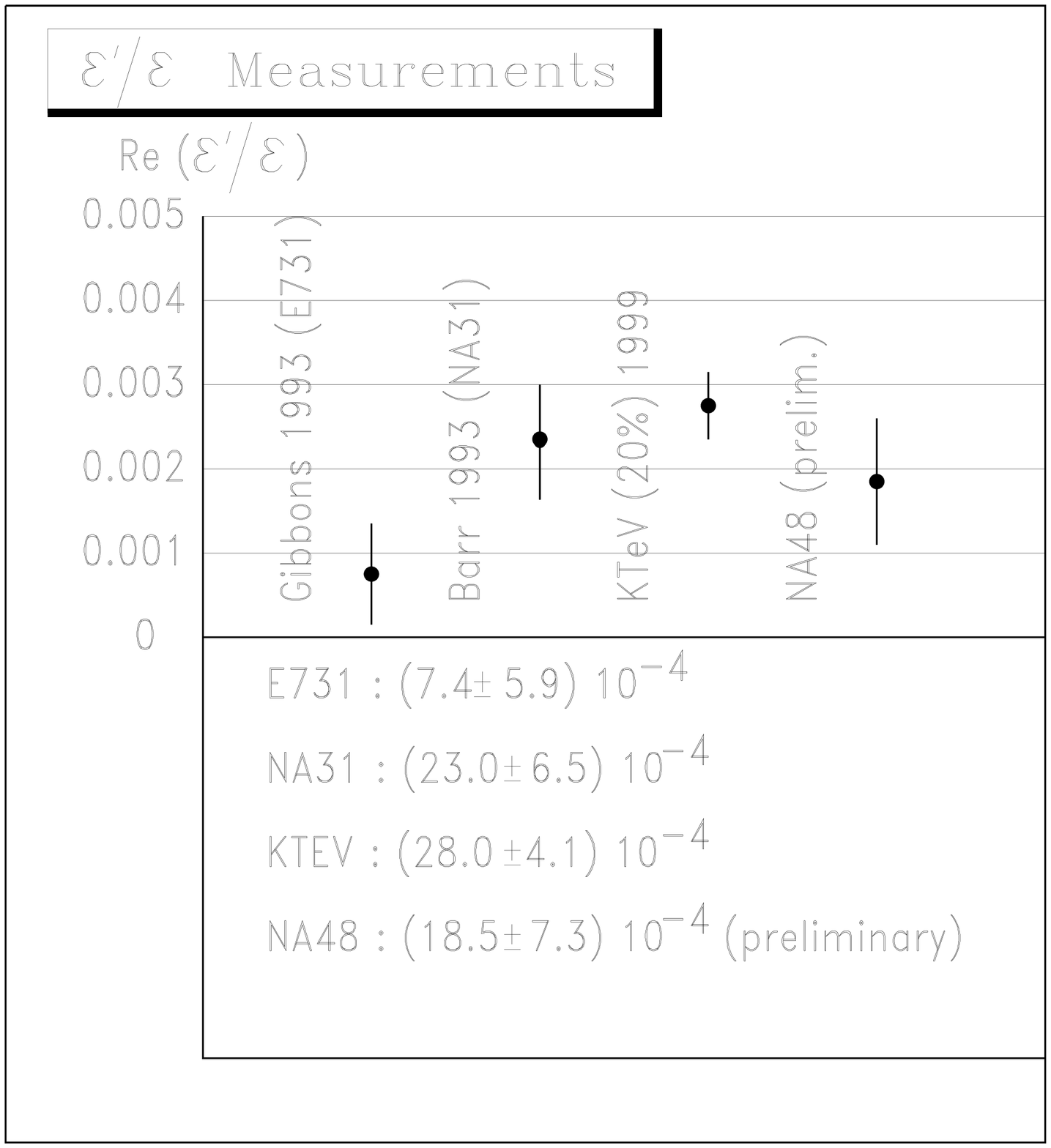}
\caption{Recent results on $ \epsilon^{\prime}/\epsilon $.}
\end{figure}

	The KTeV result is in better agreement with that from
NA31 rather than from E731 but the experimenters have not found
any reason other than a fluctuation to account for this difference.
The E731 beam, detector, and analysis were extensively documented
in a long article\footnote{CP and CPT Symmetry Tests from the
Two-Pion Decays of the Neutral Kaon with the Fermilab E731
Detector with L.K. Gibbons et al., Physics Review \textbf{D55},
6625 (1997)}
for Physical Review D.  At present the NA48 result is in good 
agreement with those from E731, KTeV, and NA31.

	The theoretical situation is still developing.  Figure 9
shows this grand average along with some recent predictions.  It
is still too early to say if the rather large value points to
new physics or to inadequacies in the standard model calculations.
The goals of the next round of lattice calculations, aiming
at 10\% determinations of the matrix elements, would mean that
$ \epsilon^{\prime}/\epsilon $ would become a ``powerful precision
test and new physics probe\footnote{Bill Marciano, summary
talk of the Chicago Kaon Conference, June 1999, editors
Jon Rosner
and Bruce Winstein, to be published by the University of Chicago Press.}.''

\begin{figure}[!ht]
\centering
\epsfysize=3.5in   
\hspace*{0in}
\epsffile{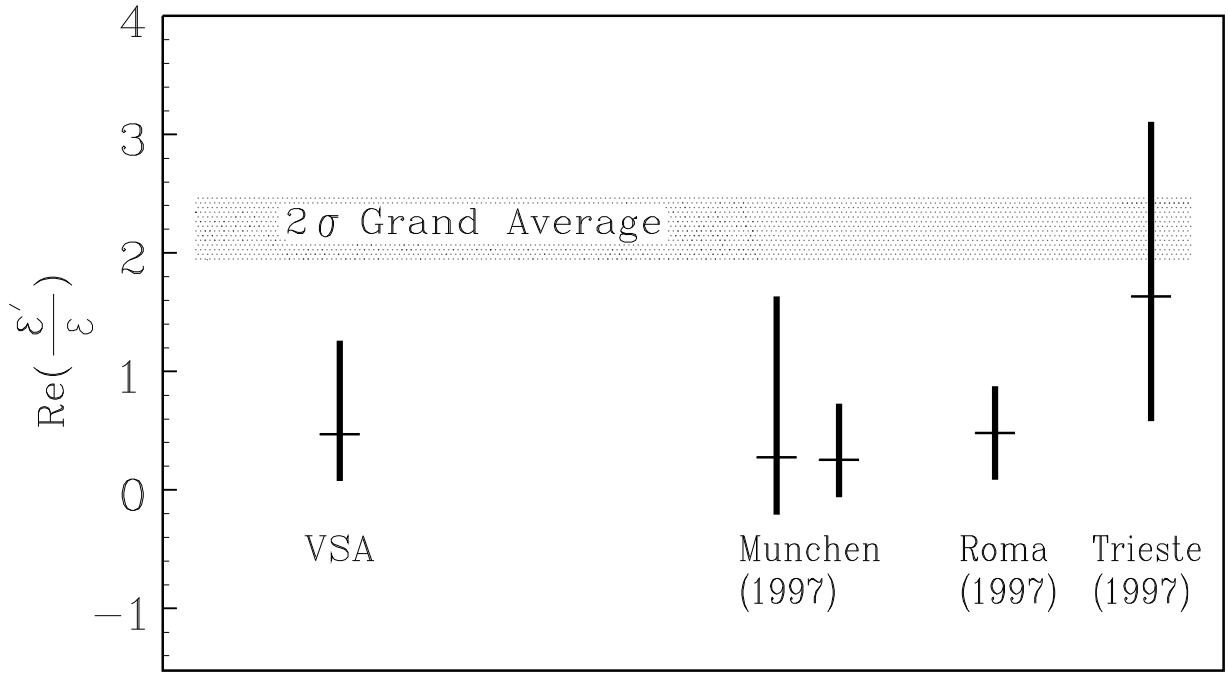}
\caption{Recent CKM estimates for Re($\epsilon^{\prime}/\epsilon $)
together with the grand average of recent experiments.}
\end{figure}

\section{RARE DECAYS} 

	In this section, I will briefly mention other studies
of rare K decays that have an implication for CP violation.

\subsection{$K_L \rightarrow \pi^0 ll $}

	A theoretically clean channel\footnote{L. Littenberg, Phys. Rev.
\textbf{D39}, 3322 (1989).}
is the decay of the long-lived Kaon to a neutral
pion, neutrino, anti-neutrino pair.  This is mediated by the
Z-penguin, with a branching ratio given by\footnote{B. Winstein, and L.
Wolfenstein, Rev. Mod. Phys. \textbf{65}, 1113 (1993).}:

\[ BR (K_L \rightarrow \pi^0 \nu \bar{\nu}) = 8 \times 10^{-11}
(M_t /M_W )^{2.2} A^4 \eta^2 \]

where $ \eta $ is the parameter in the CKM matrix.

	This mode has been the by-product of a number of
searches by Fermilab experiments.  The latest result comes from the
KTeV experiment which reports\footnote{J. Adams et al., hep-ex/9806007.}:

\[ BR (K_L \rightarrow \pi^0 \nu \bar{\nu} ) < 5.9 \times 10^{-7}
(KTeV, 90\%~\textrm{confidence}). \]

	Less clean but still interesting is the ee mode.  The KTeV
90\% confidence value is 6.64 $ \times 10^{-10} $ and the similar
limit\footnote{J. Whitmore,
Chicago Kaon Conference, ibid.}
for the
$ \mu \mu $ mode is 3.4 $ \times 10^{-10} $.  These
are both an order of magnitude improved over the corresponding E731
results and while not quite at the level expected in the Standard
Model, are beginning to rule out certain parameter regions in an
extended SUSY scheme.\footnote{Supersymmetric contributions to rare kaon decays:
beyond the single mass-insertion approximation, G. Colangelo and G.
Isidori, hep-ph/9808487.}

\section{FUTURE RARE DECAY STUDIES}

	The prime goal, now that direct CP violation has been
seen, in the K system is the study of the $ \pi^0 \nu \nu $ decay
and there are several groups seriously considering
this mode.  Many institutions in KTeV have written an expression
of interest to
pursue this mode at Fermilab, using the much higher intensity of the Main
Injector.  There is a similar proposal at BNL.  The goal is the
collection of about 100 events.  The idea is that with this mode
and the corresponding $ K^+ $ decay, one can accurately determine
the angle $ \beta $ of the unitarity triangle in the
K sector and this can be compared to a similar determination in the
B sector.

	The E787 experiment at Brookhaven has the best information
on the $ K^+ $ decay: they have seen one unambiguous event which
corresponded to a branching ratio\footnote{George Redlinger,
Chicago Kaon Conference, ibid.} in
the range of about 2 $ \times 10^{-10} $.  They have a proposal to
increase their sensitivity to allow about 10 events to be collected.
And the CKM proposal at Fermilab would collect about 100 such events.

	These experiments, if successful, would allow a test at
the level of a few degrees and thus will be quite sensitive to any
new physics.

\section{CONCLUSION}

	KTeV has reported on just 20\% of the data collected in 1996/7.
And more data is being collected now in the 1999 run, with a variety
of systematic checks, so that the accuracy on
$ \epsilon^{\prime}/\epsilon $ should improve significantly.  

	NA48 has also collected more data and should run as well in 2000.
And KLOE is collecting its first samples now.  So, during the next
few years, we should see several high precision determinations of
Re($ \epsilon^{\prime}/\epsilon $).  The theorists are actively
looking to improve their predictive power so that we should know on
that time scale if we are seeing new physics.  One thing is certain:
a new CP violating effect has been established.  It will be
quite interesting to see if such an effect can be isolated in B
decays and in any case new CP violating signatures in that sector, in
comparison with studies to be made in the kaon decays, will
tell us if the CKM picture is complete, or if new mechanisms are needed.

\section{ACKNOWLEDGEMENT}

	The author would like to acknowledge the conference organizers
for setting up a most stimulating environment, and Kathy Visak for help
in the preparation of this manuscript.

\end{document}